\documentclass[sigconf]{acmart}

\AtBeginDocument{%
  \providecommand\BibTeX{{%
    \normalfont B\kern-0.5em{\scshape i\kern-0.25em b}\kern-0.8em\TeX}}}

\setcopyright{rightsretained}
\copyrightyear{2018}

\acmConference{\vspace{2pt}SIGGRAPH 2021}{August 9--13, 2021}{Virtual}

\begin{document}

\title[Silent Speech and Emotion Recognition]{Silent Speech and Emotion Recognition from Vocal Tract Shape Dynamics in Real-Time MRI}

\author{Laxmi Pandey}
\affiliation{%
  \institution{Human-Computer Interaction Group}
  \institution{University of California, Merced}
  \streetaddress{5200 N. Lake Road}
  \state{California}
  \city{Merced}
  \country{USA}}
  \postcode{95343}
\email{lpandey@ucmerced.edu}


\author{Ahmed Sabbir Arif}
\affiliation{%
  \institution{Human-Computer Interaction Group}
  \institution{University of California, Merced}
  \streetaddress{5200 N. Lake Road}
  \state{California}
  \city{Merced}
  \country{USA}}
  \postcode{95343}
\email{asarif@ucmerced.edu}

\begin{abstract}
Speech sounds of spoken language are obtained by varying configuration of the articulators surrounding the vocal tract. They contain abundant information that can be utilized to better understand the underlying mechanism of human speech production. We propose a novel deep neural network-based learning framework that understands acoustic information in the variable-length sequence of vocal tract shaping during speech production, captured by real-time magnetic resonance imaging (rtMRI), and translate it into text. The proposed framework comprises of spatiotemporal convolutions, a recurrent network, and the connectionist temporal classification loss, trained entirely end-to-end. On the USC-TIMIT corpus, the model achieved a 40.6\% PER at sentence-level, much better compared to the existing models. To the best of our knowledge, this is the first study that demonstrates the recognition of entire spoken sentence based on an individual's articulatory motions captured by rtMRI video. We also performed an analysis of variations in the geometry of articulation in each sub-regions of the vocal tract (i.e., pharyngeal, velar and dorsal, hard palate, labial constriction region) with respect to different emotions and genders. Results suggest that each sub-regions distortion is affected by both emotion and gender.
\end{abstract}


\begin{teaserfigure}
  \centering
  \includegraphics[width=\textwidth]{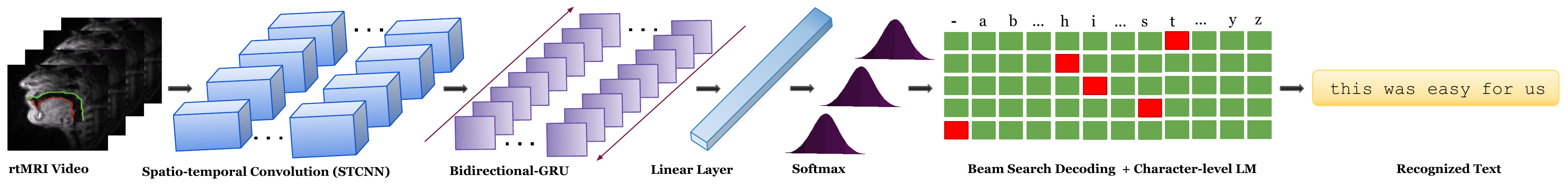}
  \caption{An overview of the proposed model:  classification of 2D real-time MRI (rtMRI) of vocal tract shaping into text with an end-to-end deep neural network. A sequence of frames is used as input that is processed by three layers of STCNN for spatiotemporal feature extraction. The extracted features are processed by two Bi-GRUs, followed by a linear layer and a softmax layer. The softmax output is then decoded with prefix beam search with the help of a language model.}
  \label{fig:teaser}
\end{teaserfigure}

\maketitle

\section{Introduction}
The vocal tract is the most important component of human speech production that starts at the vocal cords, continues upwards towards the tongue, and ends at the lips \cite{s_study_2016}. During air expulsion, this tubular passageway changes its position and shape to produce various sounds and their acoustic representations. Estimating and mapping vocal tract configuration to its corresponding acoustic parameters has long been a challenge not only because it is difficult to access the vocal tract but also due to its complex biological structure and rapid movement of its articulators \cite{mitra_joint_2017}. The speech production process is essentially non-stationary---generally the rapid transition between different articulatory states generates the speech sounds. Hence, extraction of acoustic information embedded in the vocal tract geometry is crucial for recognizing and synthesizing speech.

Recent development in real-time magnetic resonance imaging (rtMRI) makes it possible to acquire complex spatiotemporal visual information about the dynamic shaping of the vocal tract in speech production needed for speech analysis. Unlike X-ray or electromagnetic articulography (EMA), rtMRI does not use potentially hazardous radiation or place extramural devices in the mouth that could interfere with the articulator's movement. This method is also suitable for patients with vocal tract pathology, such as those that have a partial tongue resection or experience pain in the areas responsible for producing speech. The University of Southern California collected a speech production dataset\footnote{USC-TIMIT: A Database of Multimodal Speech Production Data, \url{https://sail.usc.edu/span/usc-timit}\label{USC-TIMIT}} that includes rtMRI data from ten native speakers of general American English \cite{narayanan_real-time_2014}. We utilized this dataset to investigate whether it is possible to recognize continuous speech from vocal tract geometry.

For this, we built a deep neural network-based learning framework that can automatically estimate the acoustic information corresponding to a specific vocal tract configuration, called articulatory-to-acoustic mapping, for continuous speech recognition. This, we believe, is the first end-to-end sentence-level articulatory speech recognition\footnote{Articulatory speech recognition identifies a sequence of characters based on the corresponding sequence of vocal tract shapes.} framework for rtMRI data that simultaneously learns spatiotemporal visual features and sequential information. In addition, we performed an extensive analysis on the MR images of emotion-dependent vocal tract movements to compare different emotions (neutral, happy, angry, sad) and genders (female, male) using the data\footnote{USC-EMO-MRI: An Emotional Speech Production Database, \url{https://sail.usc.edu/span/usc-emo-mri}} collected in a previous work \cite{kim_usc-emo-mri_2014}. An understanding of whether and how emotion affects articulatory movements during speech production is important to reduce ambiguity in recognized sentences. For example, the emotional context of the sentence ``I hate you'' could inform the system whether it was said sarcastically or literally. The effects of gender on vocal tract movements, in contrast, can increase the accuracy of the recognition system.

We envision numerous applications of this framework. It could be used to input text and communicate with various computer systems using speech or silent speech \cite{10.1145/3411764.3445565}, which are arguably more natural modes of interaction \cite{wigdor_brave_2011}. It can also enable users to interact with public displays and kiosks without contact \cite{10.1145/3411764.3445430}, which is of a particular interest in global spread of infectious diseases, such as the current COVID-19 situation. Most importantly, it could enable people with speech disorder, muteness, and blindness to input text and interact with various computer systems, increasing their access to these technologies.




This article starts with a review of the existing work in the area. It then explains the proposed recognition model, followed by its evaluation and comparison with previous works. It then presents an analysis of variations in the geometry of articulation in each sub-regions of the vocal tract (i.e., pharyngeal, velar dorsal, hard palate, labial constriction region) with respect to different emotion and gender. Finally, it discusses the findings and limitations of the work and concludes with speculations on future extensions.

\section{Related Work}
The feasibility of using rtMRI and deep learning to recognize speech is not well investigated in the literature. \citet{saha_towards_2018} classified vowel-consonant-vowel (VCV) combinations using the same dataset used in this work with an accuracy rate of 42\%. Leeuwen et al. \cite{van_leeuwen_cnn-based_2019} classified 27 sustained phonemes from MR images using a convolutional neural network with an accuracy rate of 57\%.

Some have also used rtMRI to study articulatory characteristics of emotional speech using vocal tract movement data \cite{kim_usc-emo-mri_2014}. Lee et al. \cite{lee_study_2006} analyzed the rtMRI data of emotional speech of one male speaker. They found out that ``angry'' speech can be characterized by much wider and faster vocal tract shaping and the extra usage of the pharyngeal region than the other examined emotions (neutral, happy, and sad). They also reported that ``happy'' speech exhibited shorter vocal tract length than the other emotions. Their findings were, however, obtained from a limited dataset collected from only one male speaker. A different work \cite{lee_vocal_2010} reported the differences in vocal tract behaviors, and between inter-speaker and intra-speaker in different speech production styles, such as different emotion expression. Kim et al. \cite{kim_usc-emo-mri_2014} found out that the pharyngeal constriction and releasing are more emphasized for ``angry'' than ``happy'', while the palatal constriction and releasing are more emphasized for ``happy'' than ``angry'' during the production of one word ``five''.

\section{Recognition Model}
The aim of our recognition model is to predict the phrase being spoken from a silent video of vocal tract movements during speech production. 
It uses the LipNet model \cite{assael_lipnet_2016} that has been used in the past to generate text conditioned on lip sequences \cite{10.1145/3411764.3445565}. However, the decoder was conditioned on vocal tract movement sequences as illustrated in Fig.~\ref{fig:teaser}.
The proposed recognition model consists of two sub-modules (or sub-networks): a \textit{feature extraction} frontend that takes a sequence of video frames and outputs one feature vector per frame, and a \textit{sequence modeling} module that inputs the sequence of per-frame feature vectors and outputs a sentence character by character. We describe these modules below.

\subsection{Feature Extraction}
The recognition model takes a sequence of $T$ frames as input to process by 3 layers of spatiotemporal convolutions (STCNN) \cite{ji_3d_2013}. It consists of a convolutional layer with 64 3-dimensional (3D) kernels of $5\times7\times7$ size (time $\times$ width $\times$ height), followed by Batch Normalization (BN) \cite{ioffe_batch_2015} and Rectified Linear Units (ReLU) \cite{agarap_deep_2019}. Each extracted feature map is passed through a spatiotemporal max-pooling layer, which drops the spatial size of the 3D feature.

\subsection{Sequence Modeling}
The extracted features are processed by 2-Bidirectional Gated Recurrent Units (Bi-GRUs) \cite{chung_empirical_2014}, where each time-step of the GRU output is processed by a linear layer, followed by a softmax layer over the vocabulary. Then,
an end-to-end model is trained with connectionist temporal classification (CTC) loss \cite{graves_connectionist_2006}. Next, the softmax output is decoded with a left-to-right beam search \cite{collobert_fully_2019} that incorporates prior information from an external language model \cite{williams_contextual_2018} to recognize the spoken utterances. The model is capable of mapping variable-length video sequences to text sequences. All layers use rectified linear unit (ReLU) activation functions \cite{agarap_deep_2019}. During inference, we use a 5-gram character-level Language Model (LM), which is a recurrent network with 4 unidirectional layers of 1,024 LSTM cells each. The LM is trained to predict one character at a time.

%

\section{Experiment}
This section describes the dataset preparation, the experiments conducted for parameter selection of the model, and the training protocol used to build the proposed model. We do parameter selection for batch size, number of epochs, and beam width ($K$). We then compare the performance of proposed articulatory speech recognition with several existing deep learning models to demonstrate that our model performs much better than those.

\subsection{Dataset Preparation}
To validate the performance of the proposed model, we performed an articulatory speech recognition experiment on the USC-TIMIT dataset$^{\ref{USC-TIMIT}}$, which includes 2D rtMRI of vocal tract shaping of ten speakers (5 female and 5 male, M = 28.7 years, SD = 7.2) along with synchronized audio recordings and their time-aligned word-level transcriptions \cite{narayanan_real-time_2014}. To prepare the labeled training data for our model, an alignment between the word-level transcription and the videos frames is needed. Hence, we estimated the number of frames by multiplying duration (second) of each word by video frame rate (23.18 frames/second) and aligned with its word-level transcription. Once we had the labeled data, we divided the total available data into a ``training dataset'' with 3,680 videos of eight speakers and a ``testing dataset'' with the remaining 920 videos of two speakers.

\subsection{Training}
Before feeding the data to the model, we augmented the training dataset by applying a horizontally mirrored transformation on video frames. In total, there were 10,972 samples. We augmented the dataset with simple transformations to reduce overfitting. We trained the model on both regular and horizontally mirrored image sequences. In addition, we varied the parameters of the model one by one keeping all others fixed, and simultaneously conducted evaluations for various combinations to select the set of optimum values. The batch size varied from 16 to 256, then set to the optimum value of 32. Similarly, the number of steps/epoch for the training was changed from 50 to 500 and fixed at 100 since it yielded a better accuracy rate. Table~\ref{table1} summarizes the hyperparameters of the recognition model, where $T$ denotes the number of frames, $H$ and $W$ denote the height and width, respectively, $C$ denotes channels, $F$ denotes the feature dimension, and $V$ denotes the number of characters in the vocabulary.

\begin{table}[h]
  \centering
  \begin{tabular}{l r r}
     \toprule
    
    {\small\textit{Layer}}
    & {\small \textit{Dimension}}
      & {\small \textit{Order}}
    \\
    \midrule
   InputLayer & 75 x 64 x 64 x 1 & T x W x H x C\\
    ZeroPadding3D & 77 x 68 x 68 x 1 & T x W x H x C  \\
    Conv3D & 75 x 32 x 32 x 32 & T x W x H x C \\
    BatchNorm & 75 x 32 x 32 x 32 & T x W x H x C \\
    Activation & 75 x 32 x 32 x 32 & T x W x H x C \\
    Dropout & 75 x 32 x 32 x 32 & T x W x H x C \\
    MaxPool3D & 75 x 16 x 16 x 32 & T x W x H x C \\
    ZeroPadding3D & 75 x 20 x 20 x 32 & T x W x H x C \\
    Conv3D & 75 x 16 x 16 x 64 & T x W x H x C \\
    BatchNorm & 75 x 16 x 16 x 64 & T x W x H x C \\
    Activation & 75 x 16 x 16 x 64 & T x W x H x C \\
    Dropout & 75 x 16 x 16 x 64 & T x W x H x C \\
    MaxPool3D & 75 x 8 x 8 x 64 & T x W x H x C \\
    ZeroPadding3D & 75 x 10 x 10 x 64 & T x W x H x C \\
    Conv3D & 75 x 8 x 8 x 96 & T x W x H x C \\
    BatchNorm & 75 x 8 x 8 x 96 & T x W x H x C \\
    Activation & 75 x 8 x 8 x 96 & T x W x H x C \\
    Dropout & 75 x 8 x 8 x 96 & T x W x H x C \\
    MaxPool3D & 75 x 4 x 4 x 96 & T x W x H x C \\
    Bi-GRU & 75 x 512 & T x F \\
    Bi-GRU & 75 x 512 & T x F \\
    Linear & 75 x 28 & T x V \\
    Softmax & 75 x 28 & T x V \\
    \bottomrule
  \end{tabular}
  \caption{Recognition model architecture hyperparameters.}~\label{table1}
\end{table}

The number of frames was fixed to 75, $\sim$3 seconds. Longer image sequences were truncated and shorter sequences were padded with zeros. We batch-normalized the outputs of each convolution layer. All layers used rectified linear unit (ReLU) activation functions. We applied a dropout \cite{srivastava_dropout_2014} of 0.5 after each max-pooling layer. The model was trained end-to-end by the Adam optimizer \cite{kingma_adam_2017} with a batch size of 32. The learning rate was set to $10^{-3}$. The Phoneme Error Rate (PER), Character Error Rate (CER), and Word Error Rate (WER) were computed using the CTC beam search (see Section \ref{evaluation}). On top of that, we used a character 5-gram binarized language model. The above-described network model was implemented with the Keras deep-learning platform with Tensorflow \cite{abadi_tensorflow_2016} as the backend, and an NVIDIA GeForce 1080Ti as the GPU board. Training network with 10,972 samples required approximately 3.5 hours.

\section{Results}
We evaluated the proposed architecture and training strategies. We also compared the model with previous work on articulatory speech recognition \cite{saha_towards_2018,van_leeuwen_cnn-based_2019} that considered only the simpler case of predicting vowel-consonant vowel (VCV) combinations and phoneme from static MR images using a deep neural network. Note that our model, in contrast, predicts sequences, thus can exploit temporal context to attain higher accuracy. The inference and evaluation procedures used in this work are described below.

\subsection{Beam Search}
\begin{figure}[h]
\centering
  \includegraphics[width=0.4\textwidth]{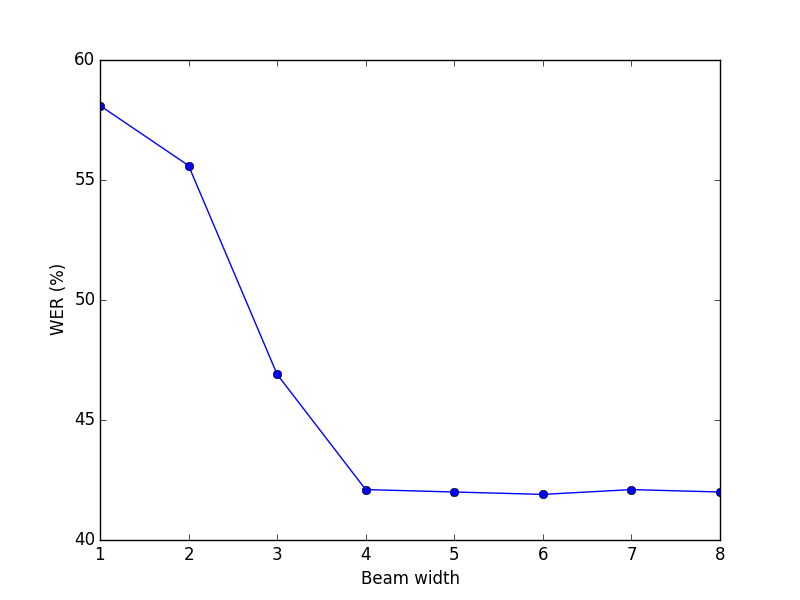}
  \caption{The effect of beam width on Word Error Rate (WER).}~\label{beam}
\end{figure}

As discussed earlier, our architecture performs character-level prediction on input frames by performing CTC beam search of width 4. At each timestep, the hypotheses in the beam are expanded with every possible character, and only the 4 most probable hypotheses are stored. Fig.~\ref{beam} illustrates the effect of increasing the beam width, where one can see that there is no observed benefit for increasing the width beyond 4.

\begin{table*}[h]
\centering
\begin{tabular}{ccccc}
\hline
\textbf{Dictionary}          & \textbf{Dataset}       & \textbf{PER \%} & \textbf{CER \%} & \textbf{WER \%} \\ \hline
Vowel-Consonant-Vowel \cite{saha_towards_2018} & Vocal Tract
Morphology MRI              &58.0   & -                           & -                   \\ \hline
Phoneme \cite{van_leeuwen_cnn-based_2019} &Vocal Tract
Morphology MRI &57.0 & -                 & -            \\ \hline
\textbf{Phrases without LM} & \textbf{USC-TIMIT} & \textbf{44.1}  & \textbf{41.7}                  & \textbf{45.4}             \\ \hline
\textbf{Phrases with LM} & \textbf{USC-TIMIT} & \textbf{40.6}  & \textbf{39.4}                  & \textbf{42.1}             \\ \hline
\end{tabular}
\caption{Performance of the three examined speech recognition models exploiting vocal tract dynamics on unseen data. The last two rows present the performance of the model proposed in this paper. Note that for a fair comparison between the models, we converted the accuracy reported in the respective papers to PER.}
\label{table2}
\end{table*}

\begin{figure*}[h]
\centering
  \includegraphics[width=\textwidth]{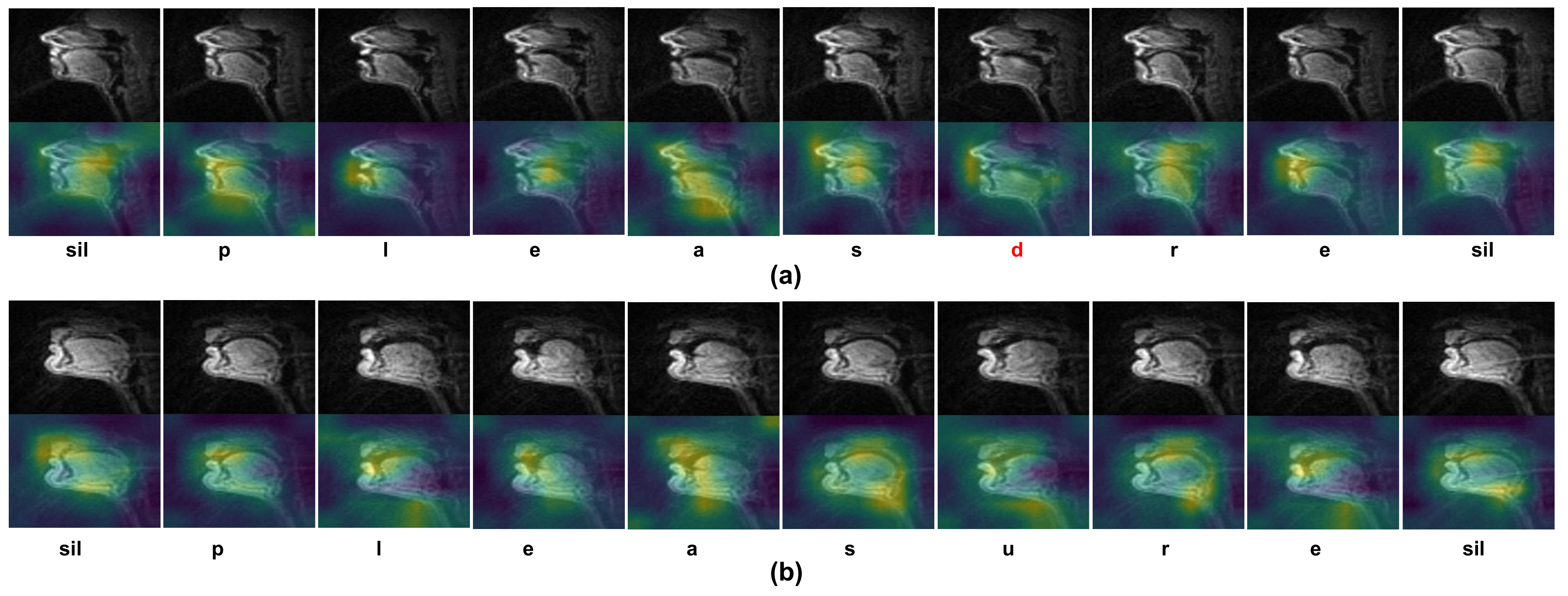}
  \vspace{-5.5mm}
  \caption{Saliency maps for the word ``pleasure'': female (a) and male (b) speakers with their corresponding phoneme predictions at the bottom. Red labels indicate incorrect predictions. Yellow shades indicate high sensitivity, that is, small changes in these pixels in the input have a large effect on the predicted class.}~\label{saliency}
\end{figure*}

\subsection{Evaluation Protocol}\label{evaluation}
To measure the performance of the proposed model, we computed the Phoneme Error Rate (PER), Character Error Rate (CER), and Word Error Rate (WER), which are standard metrics for the performance of Automatic Speech Recognition (ASR) models. A predicted word is considered correct when each character of the word is correct. PER is measured by dividing the total number of phoneme errors (the minimum number of phoneme insertions, substitutions, and deletions required to transform the predicted phrase into the ground truth) divided by the total number of phonemes. CER and WER are calculated using the same approach with phoneme replaced with character and word, respectively. Table \ref{table2} summarizes the overall PER, CER, and WER on the unseen test data. The mean PER for the proposed method is 40.6\%, which is much lower than the existing models. We also conducted an ablation study to analyze the effect of the language model on the overall performance gain. Results revealed that the model with LM exhibited 7.9\% reduction in PER, 5.5\% reduction in CER, and 7.2\% reduction in WER. It demonstrates the feasibility of the model in recognizing phrases from MRI data. The results are particularly encouraging since they suggest that even greater performance can be attained with continued exploration of this interesting and novel problem space. Besides, based on the literature, increasing the amount of training data can significantly improve recognition performance.

\subsection{Saliency}
We applied saliency visualisation techniques  \cite{simonyan_deep_2014} to interpret our model's learned behaviour, showing that it attends to important articulatory regions in the videos. Fig. \ref{saliency} illustrates analysis of two saliency visualisations for the word ``pleasure'' for female and male speakers. Notice that the regions where changes in the input have the most impact on the prediction light up. The saliency maps show that the model has learned to focus on the parts of the input frames that represent the crucial articulatory positions needed to distinguish between different phonemes. Most phonemes show a more widespread field between the tongue and palate. As can be seen in Fig.~\ref{saliency}, the saliency maps are not similar between the two subjects since vocal tract configurations varies from person to person. Notice that, in Fig.~\ref{saliency} (a), the model incorrectly predicted the phoneme \textit{d} instead of \textit{u} (highlighted in red). This mistake was made when the saliency maps showed places of attention that were not considered to be important for classification.

\section{Emotion and Gender Analysis}
\begin{figure}[htbp]
  \centering
  \begin{minipage}[b]{0.45\textwidth}
    \includegraphics[clip, trim=1cm 7cm 0.5cm 8cm, width=\textwidth]{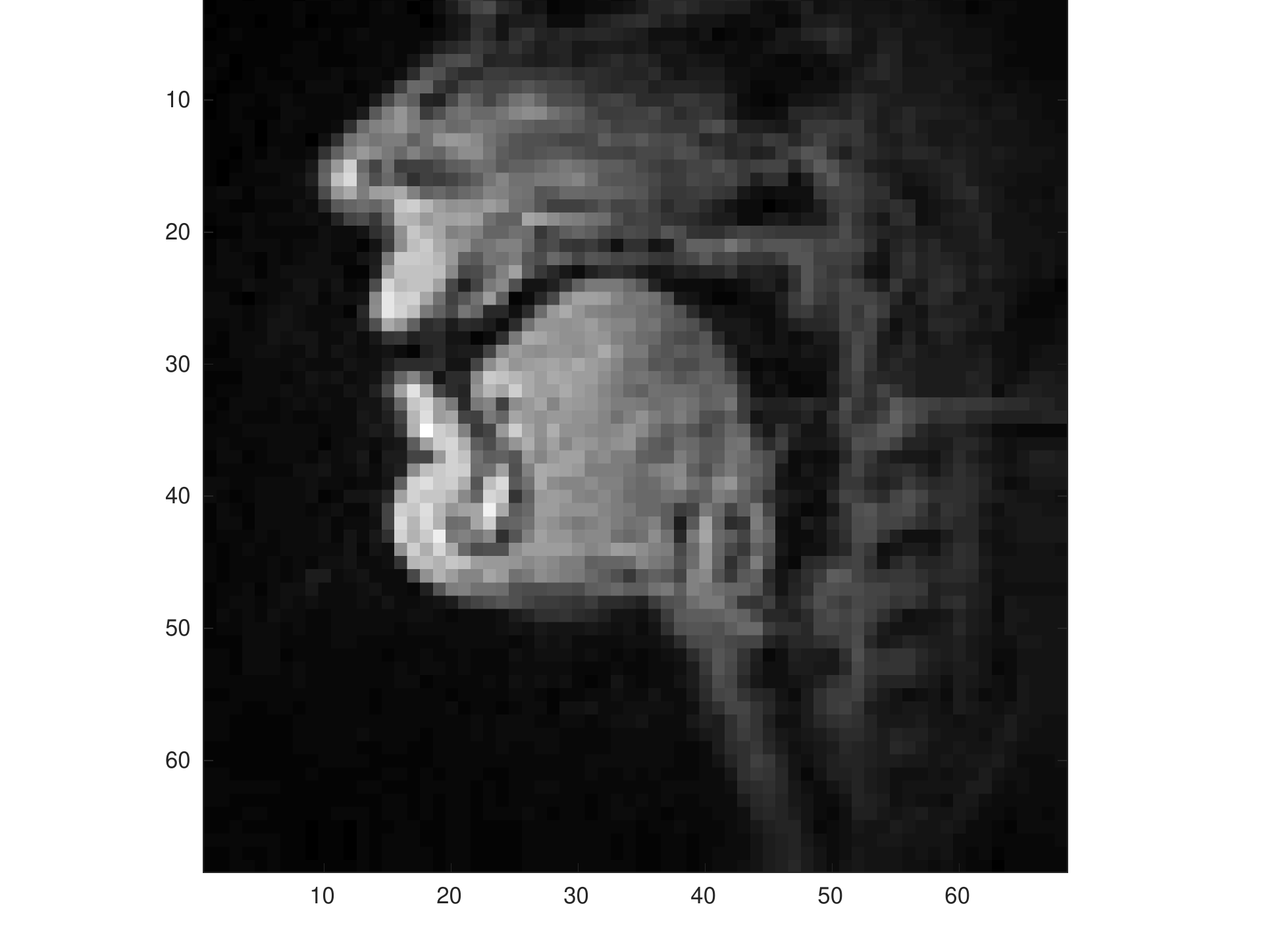}
  \end{minipage}
\hfill
  \begin{minipage}[b]{0.45\textwidth}
    \includegraphics[clip, trim=1cm 7cm 0.5cm 8cm, width=\textwidth]{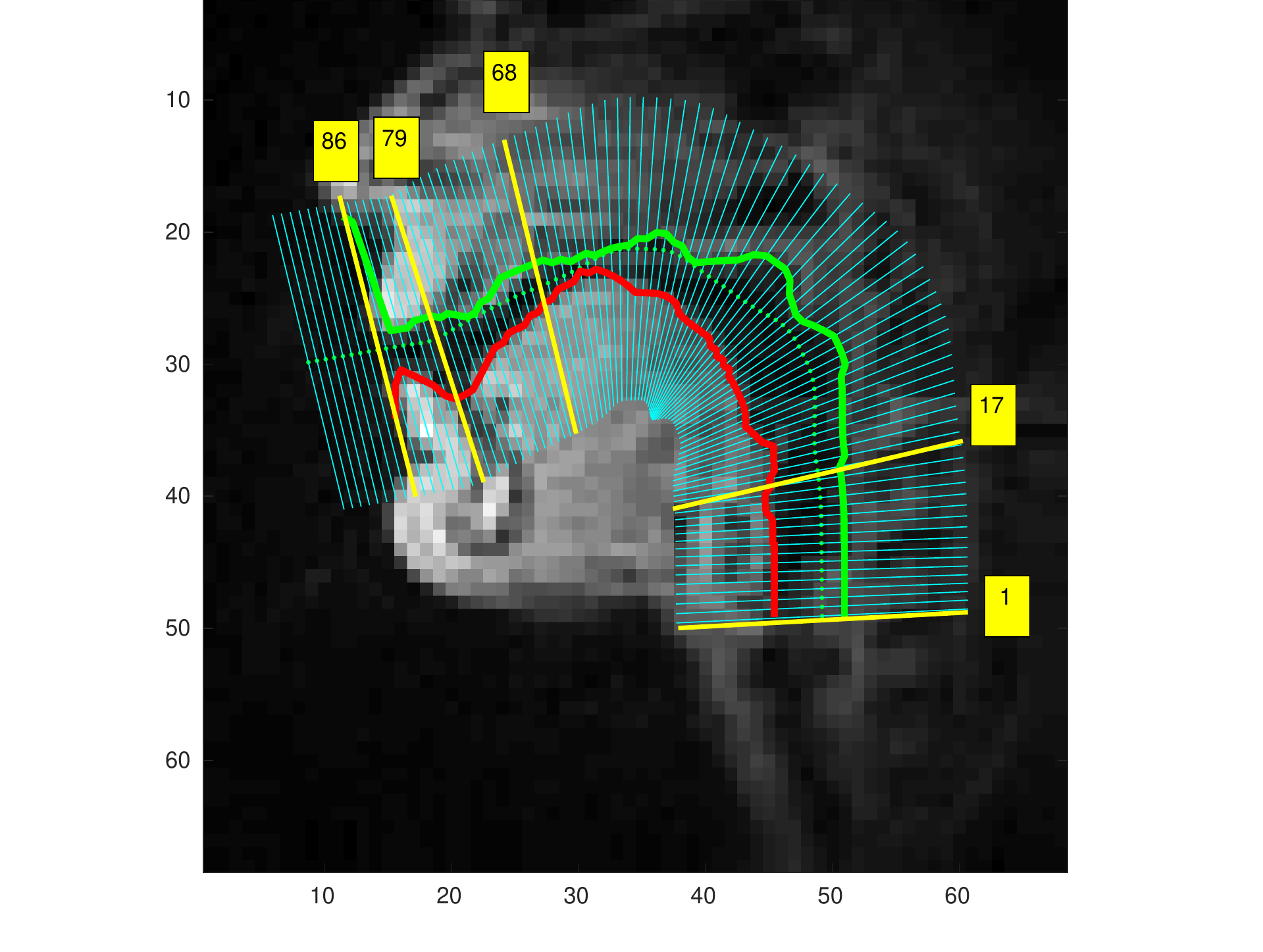}
  \end{minipage}
  \caption{Frame of neutral emotion rtMRI video showing a reference image (top) and corresponding segmentation of lower and upper boundary of vocal tract (bottom).}
  \label{mri_segment}
\end{figure}

This section presents the results of an analysis of MR images for emotion-dependent vocal tract movements to relate different emotions, particularly happy, angry, and sad, with a neutral emotion. Following the methodology used in a previous work \cite{kim_enhanced_2014}, we first extract the vocal tract airway-tissue boundaries (red line for lower boundary and green line for upper boundary in Fig.~\ref{mri_segment}), then divide them into four sub-regions: (1) grid lines 1---17 for pharyngeal region, (2) grid lines 18---68 for velar and dorsal constriction region, (3) grid lines 69---79 (alveolar ridge landmark) for the hard palate region, and (4) grid lines 80---86 for labial constriction region.

Then, we compare vocal tract shaping of different emotions by measuring the distortion in the shaping of each sub-region for each emotion ($e$) relative to neutral emotion ($n$). This is done by the normalized sum of differences of the cross-distances in the 2D space from the centroid region (mean of all the points on vocal tract airway-tissue boundaries) to each respective landmark (number of points on vocal tract airway-tissue boundaries). The cross-distances are individually computed for lower and upper boundary of each sub-region. To measure this, we developed a new metric, Neutral Emotion Deviation Measure (NEDM), defined as follows.

\begin{equation}
   \text{NEDM}_r^b = \sum_{l}\frac{|d_{n{_l}}-d_{e{_l}}|}{d_{n{_l}}}
\end{equation}

Where,

$r$: number of sub-regions in the vocal tract (i.e., 4),

$b$: lower and upper boundaries,

$l$: number of landmarks in each sub-region,


$d_{n{_l}}$: Euclidean distance between centroid  and the landmarks in the particular sub-region for neutral emotion ($n$),
\begin{align}
 d_{n{_l}} & = \textit{Euclidean-Distance}(p_{centroid},p_{n{_l}}) \nonumber \\
  & = \sqrt{\sum_{l=i}^{n}(P_{centroid} -P_{nl})^2}
\label{eq:1}
\end{align}



 $d_{e{_l}}$: Euclidean distance between centroid and the landmarks in the particular sub-region for different emotions ($e$) (i.e., happy, angry, sad),
\begin{align}
 d_{e{_l}} & = \text{Euclidean-Distance}(p_{centroid},p_{e{_l}}) \nonumber \\
  & = \sqrt{\sum_{l=i}^{n}(P_{centroid} -P_{el})^2}
\label{eq:2}
\end{align}

where,

$P_{centroid}$: $x,y$ coordinate of centroid location of sub-region,

$P_{n{_l}}$: $x,y$ coordinate of landmarks location for neutral emotion,

$P_{e{_l}}$: $x,y$ coordinate of landmarks location for different emotion.\\

In order to calculate cross-distance for each sub-region, semi-automatic tissue-airway boundary segmentation is performed using a recently introduced MATLAB software \cite{kim_enhanced_2014}. This software performs (i) tracking of the lips and the larynx, (ii) segmentation of the airway tissue boundary, (iii) pixel sensitivity correction, (iv) noise suppression on the MR image, and (v) computation of the distance function. The processes (i) and (ii) are performed automatically based on the semi-automatically constructed gridlines. This work examined the vocal tract data for the words ``clock'' and ``dock''. A total of 56 productions of each word as a function of emotion spoken by ten speakers (5 male and 5 female) were analyzed (56 productions $\times$ 2 words $\times$ 3 emotions).

Table \ref{table3} presents the identification of the most affected regions in vocal tract airway-tissue upper and lower boundaries for each emotion. The reported values are averages of the distortion measure. On average, the sub-regions of lower boundary showed a greater deviation from centroid location than upper boundary regions for all emotions. The velar and dorsal constriction region and hard palate region showed more distortion for high arousal emotions (anger and happyness) than low arousal emotion (sadness). In general, the velar and dorsal constriction region was of great importance for all emotions. For low arousal emotions, all regions tended to have less noticeable changes compared to high arousal emotions. The palatal constriction and releasing were more emphasized for happiness than for anger. Results showed that the distortion factor was also affected by gender. For all emotions, female speakers had more noticeable changes in all regions. However, labial constriction region showed very less variation across gender. For anger, female speakers had more geometrical distortion in pharyngeal and velar and dorsal constriction regions than happiness.

\begin{table*}[h]
\centering
\begin{tabular}{ccccc}

\multicolumn{5}{c}{\textbf{Lower boundary geometrical comparison of each sub-region in the vocal tract with respect to different emotions}} \\ \hline
\multicolumn{5}{c}{\textbf{Clock (Male)}} \\ \hline
\begin{tabular}[c]{@{}c@{}}Regions/\\ Emotions\end{tabular} &
  \textbf{Pharyngeal} &
  \textbf{Velar and dorsal constriction} &
  \textbf{Hard palate} &
  \textbf{Labial constriction} \\ \hline
\textbf{Happy} &
  0.67 &
  0.78 &
  0.84 &
  0.62 \\ \hline
\textbf{Angry} &
  0.85 &
  0.91 &
  0.72 &
  0.73 \\ \hline
\textbf{Sad} &
  0.36 &
  0.48 &
  0.41 &
  0.50 \\ \hline
\multicolumn{5}{c}{\textbf{Clock (Female)}} \\ \hline
\begin{tabular}[c]{@{}c@{}}Regions/\\ Emotions\end{tabular} &
  \textbf{Pharyngeal} &
  \textbf{Velar and dorsal constriction} &
  \textbf{Hard palate} &
  \textbf{Labial constriction} \\ \hline
\textbf{Happy} &
  0.71 &
  0.80 &
  0.93 &
  0.64 \\ \hline
\textbf{Angry} &
  0.89 &
  1.00 &
  0.86 &
  0.74 \\ \hline
\textbf{Sad} &
  0.41 &
  0.54 &
  0.49 &
  0.53 \\ \hline
\multicolumn{5}{c}{\textbf{Dock (Male)}} \\ \hline
\begin{tabular}[c]{@{}c@{}}Regions/\\ Emotions\end{tabular} &
  \textbf{Pharyngeal} &
  \textbf{Velar and dorsal constriction} &
  \textbf{Hard palate} &
  \textbf{Labial constriction} \\ \hline
\textbf{Happy} &
  0.68 &
  0.74 &
  0.83 &
  0.62 \\ \hline
\textbf{Angry} &
  0.83 &
  0.94 &
  0.70 &
  0.72 \\ \hline
\textbf{Sad} &
  0.32 &
  0.43 &
  0.38 &
  0.53 \\ \hline
\multicolumn{5}{c}{\textbf{Dock (Female)}} \\ \hline
\begin{tabular}[c]{@{}c@{}}Regions/\\ Emotions\end{tabular} &
  \textbf{Pharyngeal} &
  \textbf{Velar and dorsal constriction} &
  \textbf{Hard palate} &
  \textbf{Labial constriction} \\ \hline
\textbf{Happy} &
  0.75 &
  0.80 &
  0.94 &
  0.61 \\ \hline
\textbf{Angry} &
  0.91 &
  0.98 &
  0.87 &
  0.69 \\ \hline
\textbf{Sad} &
  0.43 &
  0.50 &
  0.49 &
  0.48 \\ \hline\hline
  \multicolumn{5}{c}{\textbf{Upper boundary geometrical comparison of each sub-region in the vocal tract with respect to different emotions}} \\ \hline
\multicolumn{5}{c}{\textbf{Clock (Male)}} \\ \hline
\begin{tabular}[c]{@{}c@{}}Regions/\\ Emotions\end{tabular} &
  \textbf{Pharyngeal} &
  \textbf{Velar and dorsal constriction} &
  \textbf{Hard palate} &
  \textbf{Labial constriction} \\ \hline
\textbf{Happy} &
  0.42 &
  0.56 &
  0.34 &
  0.48 \\ \hline
\textbf{Angry} &
  0.37 &
  0.44 &
  0.48 &
  0.45 \\ \hline
\textbf{Sad} &
  0.30 &
  0.41 &
  0.33 &
  0.35 \\ \hline
\multicolumn{5}{c}{\textbf{Clock (Female)}} \\ \hline
\begin{tabular}[c]{@{}c@{}}Regions/\\ Emotions\end{tabular} &
  \textbf{Pharyngeal} &
  \textbf{Velar and dorsal constriction} &
  \textbf{Hard palate} &
  \textbf{Labial constriction} \\ \hline
\textbf{Happy} &
  0.44 &
  0.48 &
  0.53 &
  0.49 \\ \hline
\textbf{Angry} &
  0.61 &
  0.54 &
  0.47 &
  0.41 \\ \hline
\textbf{Sad} &
  0.33 &
  0.42 &
  0.49 &
  0.39 \\ \hline
\multicolumn{5}{c}{\textbf{Dock (Male)}} \\ \hline
\begin{tabular}[c]{@{}c@{}}Regions/\\ Emotions\end{tabular} &
  \textbf{Pharyngeal} &
  \textbf{Velar and dorsal constriction} &
  \textbf{Hard palate} &
  \textbf{Labial constriction} \\ \hline
\textbf{Happy} &
  0.40 &
  0.43 &
  0.39 &
  0.45 \\ \hline
\textbf{Angry} &
  0.37 &
  0.54 &
  0.59 &
  0.45 \\ \hline
\textbf{Sad} &
  0.31 &
  0.44 &
  0.32 &
  0.28 \\ \hline
\multicolumn{5}{c}{\textbf{Dock (Female)}} \\ \hline
\begin{tabular}[c]{@{}c@{}}Regions/\\ Emotions\end{tabular} &
  \textbf{Pharyngeal} &
  \textbf{Velar and dorsal constriction} &
  \textbf{Hard palate} &
  \textbf{Labial constriction} \\ \hline
\textbf{Happy} &
  0.48 &
  0.49 &
  0.41 &
  0.38 \\ \hline
\textbf{Angry} &
  0.57 &
  0.52 &
  0.44 &
  0.39 \\ \hline
\textbf{Sad} &
  0.36 &
  0.48 &
  0.40 &
  0.34 \\ \hline
\end{tabular}
\vspace{1mm}
\caption{Average Neutral Emotion Deviation Measure (NEDM) indicating the relation between each subregion and each emotion across gender. The displacements are calculated using centroid position for subregion POI.}
\label{table3}
\end{table*}

\section{Discussion}
In this work, we demonstrated that end-to-end deep learning framework can automatically map sequences of vocal tract shaping to entire sentences with a phoneme error rate of 40.6\%. The proposed model performed much better than the existing models \cite{saha_towards_2018, van_leeuwen_cnn-based_2019} that either consider only a simpler case of predicting vowel-consonant-vowel (VCV) combinations with an error rate of 58\% or phonemes with an error rate of 57\% rather than phrases. The findings suggest that deep learning represents a viable tool for continuous speech recognition from rtMRI. Most importantly, our proposed model does not rely on hand-engineered spatiotemporal visual features or a separately-trained sequence model. The proposed end-to-end model also eliminates the need for segmenting videos into words before predicting a sentence. Furthermore, saliency visualisations revealed that the proposed model learns to attend phonologically important regions of the vocal tract. It provides an insight into the vocal tract regions that are most important for phoneme classification. Further analysis revealed that mistakes were more frequent when the saliency maps showed places of sensitivity that were not expected to be important for classification.


Analysis of the MR images for emotion-dependent vocal tract movements to relate three different emotions, namely happy, angry, sad, with a neutral emotion provided interesting insights into the most affected regions of vocal tract during emotional speech production. We found out that the sub-regions of vocal tract lower boundary tended to have a more noticeable change than upper boundary regions for all emotions. It also showed that variation in each sub-region was affected by gender variability. Female speakers had more geometrical distortion in pharyngeal and velar and dorsal constriction regions for negative emotions (anger) than positive emotions (happiness). Overall, for all emotions, female speaker had more noticeable changes in all regions.


The proposed rtMRI-based speech recognition system could potentially be used as medium for input and interaction with various computer systems, incorporated in day-to-day usage. This approach could also enable people with speech disorder, muteness, and blindness to input and interact with computer systems, increasing their access to these technologies. Although we do not have the technology to achieve these just yet, the findings of this work shows its potential. Furthermore, one could apply the proposed methodology to the data from people with speech disorder and compare it to the speech of people without speech disorder to find out which articulators are involved in the impairment of phoneme production. The acoustic information extracted from vocal tract's dynamics might reveal how different phonemes production mechanism are related to each other. Also, an analysis of emotion-dependent vocal tract geometry of people with speech disorder could provide new insights into variations in emotions of their speech.


\section{Conclusion}
We proposed a deep learning framework that can decode text using the cues provided by the movements of the vocal tract. On the USC-TIMIT corpus, the proposed model achieves a 40.6\% PER at sentence-level, which is much lower than the existing models. Literature on deep speech recognition suggests that this performance is likely to further improve with addtional data  \cite{amodei_deep_2016}.
Furthermore, we conducted an analysis of variations in the geometry of articulation in each sub-regions of the vocal tract with respect to different emotions and genders. Results revealed that each sub-regions distortion was affected by both gender and emotion. In the future, we will extend this work to people with various speech disorders. We will also explore different learning models and compare their performance in the defined context.
\bibliographystyle{ACM-Reference-Format}
\bibliography{main}

\end{document}